\documentclass
[aps,prd,nofootinbib,
preprint
]{revtex4-1}

\usepackage{color}
\usepackage{graphicx,bm}
\usepackage{amsmath,amssymb,mathrsfs,verbatim,subfigure}
\usepackage{ulem}
\usepackage{hyperref}

\def \be {\begin{equation}}
\def \ee {\end{equation}}

\def \eq {Eq.~}
\def \Ref {Ref.~}

\def \<{\langle}
\def \>{\rangle}
\def \+{\dagger}
\def \({\left(}
\def \){\right)}
\def \[{\left[}
\def \]{\right]}

\def \o {\omega}

\def \O {\Omega}

\def \vk {\bm{k}}

\begin{document}

\title{Charmonium moving through a strongly coupled QCD plasma: a holographic perspective}

\author{Paul~M.~Hohler}
\affiliation{
Cyclotron Institute and
Department of Physics \& Astronomy, Texas A\& M University, College Station, TX
  77843-3366, USA
}
\email{pmhohler@comp.tamu.edu.edu}

\author{Yi~Yin}
\affiliation
	{
		Department of Physics, University of Illinois,
Chicago, IL 60607-7059, USA
	}
\email{yyin3@uic.edu}

%\pacs{
%11.25.Tq,
% Gauge/string duality
%14.40.Pq,%        Heavy quarkonia
%25.75.Nq,%        Quark deconfinement, quark-gluon plasma production and phase transitions in relativistic heavy-ion collisions
%12.38.Mh%        Quark-gluon plasma in quantum chromodynamics
%}

%\keywords{}

\begin{abstract}
The properties of charmonium in a strongly coupled QCD-like plasma at finite momentum are studied.
This is based upon a `` bottom-up " holographic model which has been previously
shown to reproduce charmonium phenomenology in vacuum and give a reasonable dissociation temperature
at zero momentum. The finite momentum spectral functions are presented and found to be consistent
with recent lattice results. The in-medium dispersion relation and momentum dependence of the decay width of J/$\psi$
have also been investigated. We find no signature of a subluminal limiting velocity from the dispersion relation,
while we note that the dissociation temperature decreases with momentum faster than previous
holographic models.
Based upon the dissociation temperature, a maximum momentum for J/$\psi$
in-medium is identified and its phenomenological implications on J/$\psi$
suppression are discussed
\end{abstract}

%\arxivnumber{}

\maketitle

\section{Introduction}
\label{sec:introduction}
The work of Matsui and Satz \cite{Matsui:1986dk} introduced the idea of using
heavy quarkonium as a probe of the formation of a strongly coupled quark gluon plasma (QGP).
In the deconfined phase,
$c\bar{c}$ bound states will dissociate because of color screening
and thereby exhibit a suppression relative to the confined phase.
This suppression can depend on
momentum of the heavy quark pair and has been observed experimentally in heavy ion collisions \cite{Adare:2006ns,Adare:2008sh,Chatrchyan:2012np,Adamczyk:2012ey}.
To understand the experimentally observed suppression, it is important to understand equilibrium properties of charmonium (see, e.g., Ref.~\cite{Rapp:2008tf} for a recent review).
In this paper, we will focus
on one aspect of J/$\psi$ suppression, namely the dissociation temperature.

In field theory,
the basic quantity encoding the equilibrium properties of charmonium  is the retarded Green's function $G^{\text{R}}_{\mu\nu}(\o , \vk)\sim \<J_\mu J_\nu\>$ or its spectral function $\rho (\o, \vk)$.
Here
$J^{\mu}=\bar{c}\gamma^{\mu}c$ is the conserved heavy quark vector current operator.
The dissociation of J/$\psi$ can be attributed to the disappearance of the J/$\psi$ peak in $\rho(\o,\vk)$.
In terms of the Green's function, the J/$\psi$ peak is associated with a pole , $\Omega_{J/\psi}$,
of $G^{\text{R}}_{\mu\nu}(\o, \vk)$
in the lower half of the complex $\o$ plane\footnote{
	Strictly speaking, due to the symmetry
    $(G^{\text{R}}_{\mu\nu}(\o, \vk))^{*} = G^{\text{R}}_{\mu\nu}(-\o^{*}, \vk)$,
    if $\O_{J/\psi}$ is a pole of $G^{\text{R}}_{\mu\nu}(\o, \vk)$, so is $-\O^{*}_{J/\psi}$.
  For definiteness, we specify $\O_{J/\psi}$ as the one whose real part is positive.
 }.
The real part of this pole is the in-medium mass of J/$\psi$
and determines the equilibrium abundance of J/$\psi$ mesons,
while the (minus) imaginary part of that pole
is associated with the peak width,
characterizing the interactions between J/$\psi$ and the medium.
Examining the pole structure of
$G^{\text{R}}_{\mu\nu}(\o, \vk)$ is the equivalent to examining the peak structure of $\rho (\o, \vk)$.

Due to strong coupling,
the calculation of $G^{\text{R}}_{\mu\nu}(\o, \vk)$ in QCD at
temperatures relevant to experiments is a challenging
task.
Recently,
there have been effective field theory computations of quarkonia in the
quark gluon plasma (see for example \cite{Escobedo:2011ie,Escobedo:2013tca,Brambilla:2013dpa}), though these computations tend to be more
applicable to the case of bottomonium rather than that of charmonium.
Properties of heavy quarkonium in the medium have also been studied
from first principle lattice QCD simulations,
both at rest \cite{Jakovac:2006sf, Aarts:2007pk, Ding:2012sp} and at finite momentum
\cite{PhysRevD.69.094507,Oktay:2010tf,Nonaka:2011zz, Ding:2012pt}.
Similar calculations have also been preformed for bottomonium on the lattice \cite{Aarts:2012ka}.
However, being restricted to the finite interval of
Euclidean time,
determining real time correlator $G^{\text{R}}_{\mu\nu}(\o, \vk)$
requires non-trivial analytic continuation to real space.
Thereby, results for
\textit{momentum-dependent}
real time correlators from other QCD-like strongly coupled thermal systems would be helpful.

AdS/CFT correspondence or gauge/gravity duality~\cite{Maldacena:1997re, *Gubser:1998bc, *Witten:1998qj}
provides a new theoretical tool for modeling strongly-coupled thermal media.
In particular,
various properties of QCD, both in vacuum and at finite temperature,
have been modeled within that framework
(for a review, see Ref.~\cite{CasalderreySolana:2011us} and references therein).
These techniques have been applied to charmonium moving in medium previously, though they were
typically from a ``top-down" approach \cite{Liu:2006nn,Mateos:2006nu, Ejaz:2007hg}
\footnote
{
 See also Ref.~\cite{Albash:2006ew} for a similar ``top-down'' approach for mesons in general and
Ref.~\cite{Fujita:2009wc,*Fujita:2009ca} for a comparable ``bottom-up"
description.
}.
Their results can be characterized by observing a decrease in the dissociation temperature
with momentum and dispersion relation which exhibits a subluminal group velocity at asymptotically large momentum $|\vk|$.
However, as argued elsewhere \cite{Grigoryan:2010pj}, these models do not reproduce the vacuum charmonium phenomenology exceptionally well,
particularly with the excited states. Therefore, it is useful to reexamine this problem with a
more realistic holographic model.

In this paper, we present results on in-medium properties of
moving charmonium
in both transverse and longitudinal channel
from the phenomenological (``bottom-up") holography model constructed in
Ref.~\cite{Grigoryan:2010pj}.
Compared with previous holographic models of heavy quarkonia
(see, for example, Refs.~\cite{Mateos:2006nu, Ejaz:2007hg, Fujita:2009wc,*Fujita:2009ca}),
this model has the advantage of capturing QCD charmonium spectral data in vacuum.
Therefore, it is an ideal tool to explore the properties of charmonium at finite temperature
and finite momentum.

This paper is organized in the following manner. In Sec.~\ref{sec:dipmodel}, we review the vacuum holographic
model of Ref.~\cite{Grigoryan:2010pj}.
In Sec.~\ref{sec:correlators}, this is extended to finite temperature and momentum.
We present our results in Sec.~\ref{sec:jpsi} and discuss them in Sec.~\ref{sec:discussion}.
 Finally, in Sec.~\ref{sec:summary}, we conclude.

 \section{
   \label{sec:dipmodel}
   A review of the holographic model of charmonium
 }
 In this section, we review some of the key aspects of the vacuum
 ``bottom-up" holographic  model  of  charmonium  constructed in Ref.~\cite{Grigoryan:2010pj},
and extend it to finite momentum.
According to the holographic dictionary,
the conserved heavy quark vector current operator $J^{\mu}=\bar{c}\gamma^{\mu}c$
is dual to a $U(1)$ gauge field $V_{\mu}$ on the five-dimensional asymptotic $AdS_5$ background.
The relevant part of the bulk action has the usual Maxwell form
\cite{Erlich:2005qh,Karch:2006pv}:
\begin{equation}
\label{eq:action}
S=-\frac{1}{4g^2_5}\int d^5x\sqrt{g}e^{-\phi}V_{MN}V^{MN}
\end{equation}
where $ g^2_5 $ is the 5D gauge coupling, $\phi$ is the background scalar
field which, in general, is a combination of dilaton and/or tachyon
fields, corresponding to the conformal and/or chiral symmetry breaking,
and $V_{MN}=\partial_MV_N-\partial_NV_M$.
The most general metric (up to coordinate transformations) possessing three-dimensional (3D) Euclidean isometry reads
\begin{equation}\label{eq:metric0}
ds^2=e^{2A(z)}\left(dt^2-d\bm{x}^2-dz^2\right),
\end{equation}
where $z$ denotes the coordinates in the fifth dimension. The gauge field can be factorized such that
\begin{equation}
V_M(z,\bm{x},t) = V_M(z) e^{i \vk\cdot\bm{x} - i\o t}\, .
\end{equation}
%where  the charmonium 4-momentum is $k^{\mu}=(\o, \vk)$.
The equation of motion in $V_5=0$ gauge reads
\begin{equation} \label{eq:eom-V}
\partial_z\[\,e^{B\left(z\right)} \partial_z V\,\] +
(\o^2 -\vk^2)\, e^{B\left(z\right)}V = 0 \ ,
\end{equation}
where $V$ is any of the three spatial components of $V_\mu(z)$,
%and $k^\mu k_{\mu} = \omega^2 - \vec{k}^2$,
and $B(z) = A(z) - \Phi(z)$.
Near the boundary $z\to 0$,
$B(z)$ should behavior as $B(z)\to -\log z$ to represent asymptotic $AdS_5$ geometry.
Via a Liouville transformation
\begin{equation}
  \label{eq:Liouville-0}
  \Psi = e^{B(z)/2} V \, ,
\end{equation}
Equation.~(\ref{eq:eom-V}) can be brought into the form of a
Schr\"odinger equation:
\begin{equation}
  \label{eq:Schroedinger-0}
 - \frac{d^2 \Psi}{dz^2} + U(z)\Psi = (\o^2 -\vk^2)\Psi\, ,
\end{equation}
with the ``holographic potential" given by
\begin{equation}
  \label{eq:U-B-0}
  U(z) =
\frac{B''(z)}2 + \left(\frac{B'(z)}{2}\right)^2\, .
\end{equation}
Here and hereafter a prime denotes the derivative with respect to $z$.

For the case when charmonium is at rest, $\vk =0$,
discrete values of $\omega^2 = m_n^2$, for which the Eq.~(\ref{eq:Schroedinger-0}) possesses normalized bound states $\Psi _n$, correspond to the masses $m_n$
of the charmonium states, $n=1,2,\ldots = J/\psi,\psi',\ldots\ $.
Holographic correspondence also
relates decay constants of  the charmonium states to the (second) derivative of the normalized bound states $\Psi _n(z)$
(see, for example, Refs.~
\cite{Son:2003et,Erlich:2005qh}):
\begin{equation}
  \label{eq:decay-consts}
f_n =  \frac1{g_5m_n}\, {\(\sqrt z\,\Psi_n(z)\)''}\bigg|_{z\to 0}\, .
\end{equation}
In the spirit of the ``bottom-up" approach,
this model chooses the function $B(z)$
so as to satisfy the spectroscopic data associated with
charmonium.
It is assumed that such a background arises
dynamically,
but no attempt to model the corresponding
dynamics is made.
To this end, $B(z)$ is chosen such that
holographic potential, Eq.~\eqref{eq:U-B-0}, reads:
\begin{equation}
    \label{eq:piecewise-U}
    U(z) = \frac3{4z^2}\,\theta(z_d-z) +
    \((a^2 z)^2 + c^2 \)\theta(z-z_d)- \alpha\,\delta(z-z_d)\, .
\end{equation}
The holographic potential in Eq.~\eqref{eq:piecewise-U} uses
a
soft-wall model, \cite{Karch:2006pv}, for large $z$ which successfully describes light mesons, and adds
to it a ``shift term,"  $c^2$, which accounts for the mass of J/$\psi$ and a delta function
``dip" to account for the relatively large J/$\psi$ decay constant.
This model will be henceforth referred to as the ``shift and dip" model.
In Ref.~\cite{Grigoryan:2010pj}, the
four parameters in this potential have been determined by requiring that the model
correctly reproduce the masses and decay constants of J/$\psi$ and $\psi'$ yielding
\begin{equation}\label{eq:parameters}
  \begin{split}
    &a=0.970\mbox{ GeV},\  c=2.781\mbox{ GeV},\\
    &\alpha=1.876\mbox{ GeV},\ z^{-1}_d=2.211\mbox{ GeV}\, .
  \end{split}
\end{equation}
With this choice for $U(z)$, the function $B(z)$ can be determined from Eq.~\eqref{eq:U-B-0}.
For non-zero momentum, $\vk \neq 0$, the gauge field can be determined by solving
Eq.~(\ref{eq:Schroedinger-0}) with the same holographic potential $U(z)$ with identical parameters as the case when charmonium is at rest.

Prior to the ``shift and dip" model,
heavy quarkonia had been studied in the framework of gauge/gravity duality from both ``top-down" approach \cite{Mateos:2006nu,Ejaz:2007hg}
and ``bottom-up" approach \cite{Fujita:2009wc,*Fujita:2009ca}.
One crucial difference between holographic models mentioned above and the ``shift and dip" model is the spectrum of heavy quarkonium.
In both classes of models previously considered, there is only one scale which sets the masses of both the ground state
and the excited states, thus $m_n \sim n m_1$, $n \geq 1$.
In contrast, in the heavy quark limit of QCD,
the mass of the ground state, J/$\psi$,
is controlled by one parameter (the heavy quark
mass) which is different from the parameter (string tension, or $\Lambda_{\rm
  QCD}$) controlling the level spacing of excited states: $m_n^2\sim
m_1^2+(n-1)\,\Lambda_{\rm QCD}^2$.
Roughly speaking,
these two scales are represented in the ``shift and dip" model by the
parameters $c$ and $a$, respectively.

It is reasonable to expect that reproducing the charmonium spectral data, such as masses and decay constants in vacuum,
is an important prerequisite for modeling behavior of charmonium at finite temperature.
Therefore we choose the ``shift and dip" model,
which mimics QCD charmonium at zero temperature,
as a basic tool to explore in-medium properties of charmonium.
As a non-trivial test of the ``shift and dip" model,
one of us (P.H.) has applied the heavy-quark QCD sum rules to this model \cite{Hohler:2010hc}.
There is strong agreement between the moments of the polarization function calculated from the ``shift and dip"  model
and the experimental data,
suggesting
that the ``shift and dip"  model  is consistent with the heavy-quark QCD sum rules at zero temperature.
Explicit consistency checks of finite temperature QCD sum rules have not been made.

\section{Thermal correlators from holography
\label{sec:correlators} }
At non-zero temperatures, gauge plus rotation invariance requires that the retarded Green's function
$G^{\text{R}}_{\mu\nu}(\o , \vk)$
has the form
\begin{equation}
   G^{\text{R}}_{\mu\nu}(\o , \vk) =
                   P_{\mu\nu}^T(\o, \vk) \, \Pi^T(\o,\vk) +
                   P_{\mu\nu}^{L}(\o, \vk) \, \Pi^L(\o, \vk)\,,
\end{equation}
for four momentum $k^{\mu} = (\o, \vk)$.
The transverse and the longitudinal projectors are
defined in the standard way as
$P_{00}^T=0$, $P_{0i}^T=0$, $P_{ij}^T= \delta_{ij} - k_i k_j/k^2$,
and $P_{\mu\nu}^L \equiv k_\mu k_{\nu}/(k^{\mu}k_{\mu)}-g_{\mu\nu} - P_{\mu\nu}^T$, where
$i,j$ are spatial indices and $k \equiv |\vk|$.
$\Pi^{T}(\o, k), \Pi^{L}(\o, k)$ are the transverse and longitudinal correlators respectively.
In the present paper,
we are using the mostly minus signature as indicated by Eq.~\eqref{eq:metric0}.
The imaginary parts of $\Pi _{T, L}$ are related, by definition,
to transverse and longitudinal spectral functions:
\begin{equation}
\rho_{T, L}(\o, k) = - \text{Im} \Pi _{T, L} (\o, k).
\end{equation}

To study  $\rho_{T,L}(\o, k)$ from holography at finite temperature,
a blackhole ansatz for the metric is chosen:
\begin{equation}\label{eq:metricT}
ds^2=e^{2A(z)}\left(h dt^2-d\bm{x}^2- h^{-1}dz^2\right).
\end{equation}
If the function $h(z)$ has a simple zero, {\it viz.}
$h (z_h)=0$, the space described by
(\ref{eq:metricT}) possesses a horizon at $z=z_h$.
The temperature $T$ corresponding to this background is related to $z_h$ as
\begin{equation}
  \label{eq:T-zh}
  T = \frac1{4\pi} \left|h '(z_h)\right|\ .
\end{equation}
We assume the
simplest ansatz for $h$, namely,
\begin{equation}
\label{eq:f-z4}
h(z)=1-(z/z_h)^4\, ,
\end{equation}
which turns out to be the same as that in the familiar $AdS_5$ blackhole solution.
The temperature $T$ corresponding to this background is related to $z_h$
as $ z_h=(\pi T)^{-1}$.
In principle, we should expect the function
$A(z)$ as well as the background $\Phi(z)$ to depend on the
temperature.
Since we do not attempt to model the background
dynamically even at zero temperature,
we will use the simplest form of temperature dependence
as it is conventionally assumed when studying temperature-dependent quantities in ``bottom-up" models
(see e.g.~Refs.~\cite{Herzog:2006ra, Grigoryan:2010pj,Fujita:2009wc,*Fujita:2009ca}).

With the metric in Eq.~\eqref{eq:metricT}, the
transverse
and longitudinal fluctuations, $V_T$ and $V_L$, at finite momentum $k$ satisfy the equations:
\begin{subequations}
\label{eq:V}
\begin{equation}
\label{eq:VT}
V''_T(z) +\(\, B'(z) + \frac{h'(z)}{h(z)}\, \)V'_T(z) + \frac{\omega^2 - k^2 h(z)}{h^2(z)}V_T(z)
= 0\, ,
\end{equation}
\begin{equation}
\label{eq:VL}
V''_L(z) +\(\, B'(z) + \frac{\omega^2}{(\omega^2 - k^2 h(z))}\frac{h'(z)}{h(z)}\, \)V'_L(z)
+ \frac{\omega^2 - k^2 h(z)}{h^2(z)}V_L(z)
= 0\, .
\end{equation}
\end{subequations}
At zero momentum, Eqs.~\eqref{eq:VT} and \eqref{eq:VL} are identical as expected.
$\Pi_{T, L}(\o, k)$
can be calculated according to standard holographic prescriptions
\cite{Son:2002sd,*Herzog:2002pc}:
\begin{equation}
  \label{eq:G_R-V}
  \Pi _{T, L}(\omega, k ) = -\frac{1}{g_5^{2}}\, h e^B\frac{V_{T, L}'(z,\omega, k)}{V_{T, L}(z,\omega, k)}
  \bigg|_{z=\epsilon}=
  -\frac1{g_5^{2}}\,\frac{V_{T, L}'(\epsilon,\o, k)}{\epsilon V_{T, L}(\epsilon,\omega, k)},
\end{equation}
where $\epsilon \to 0$ is an ultraviolet regulator and
$V(z,\o, k)$ is the solution to Eq.~\eqref{eq:V}
satisfying infalling wave boundary conditions near the horizon:
\begin{equation}
  \label{eq:bc-V}
V(z , \omega, k )\xrightarrow{z\to z_h}
   (1-z/z_h)^{-i\omega/(4\pi T)}\, .
 \end{equation}
Equation \eqref{eq:V} with the boundary condition of \eq\eqref{eq:bc-V} must be solved numerically in the present model.
The longitudinal equation Eq.~\eqref{eq:VL} is slightly more subtle
as it will have an integrable singularity
if $\o^2 - k^2 h(z)$ has a root between the boundary and the horizon for fixed $\o$ and $k$.
However, this singularity may be avoided by making an infinitesimal Wick rotation as pointed out in Ref.~\cite{CaronHuot:2006te}.

At finite temperature,
it is also useful to apply the Liouville transformation
\begin{equation}
  \label{eq:Liouville-T}
  \Psi(\xi(z)) = e^{B(z)/2} V_T(z)\, ,
  \qquad
  \xi(z) = \int ^{z}_0 \frac{d z'}{h(z')}
\end{equation}
to the transverse equation of motion Eq.~(\ref{eq:VT}).
In the new coordinate $\xi$, we arrive at a ``Schr\"odinger" equation:
\begin{equation}
  \label{eq:Schroedinger-T}
 \frac{d^2\Psi}{d\xi^2} + \[(\o^2 - k^2)- U_{\text{sch}}(\xi , k) \]\Psi = 0\, ,
\end{equation}
with the finite temperature ``holographic potential" given by
\begin{equation}
  \label{eq:U-B-T}
  U_{\text{sch}}(\xi(z), k) =
h(z)^2\[ \frac{B''(z)}{2} + \left(\frac{B'(z)}{2}\right)^2 + \frac{B'(z)h'(z)}{2  h(z)}\] - k^2 z^4\, .
\end{equation}
A similar prescription is not as useful for the longitudinal equation of motion because the
resulting Schr\"{o}dinger coordinate
and Schr\"{o}dinger potential are in general
complex.

\section{J/$\psi$ moving in a strongly coupled plasma \label{sec:jpsi}}

\subsection{Rescaled spectral functions \label{sec:rho} }
It is convenient to define a dimensionless rescaled spectral density \cite{Grigoryan:2010pj}:
\begin{equation}
  \label{eq:rho-bar}
  \bar\rho_{T,L}(\omega, k)\equiv
\frac{\[\rho_{T,L}(\o , k)/\o^2\]}
{
	\[\rho_{T,L}(\o, 0)/\o^2\]|_{\o\to\infty}
}
=
\frac{2g_5^2}{\pi}\frac{\rho (\o, k)}{\omega^2}
\end{equation}
such that $\bar\rho_{T,L}(\o \to \infty, k)=1$.
Here we have used the fact that in large $\o$
limit,
$\rho_{T,L}(\o, k )/\o^2$ will approach its zero temperature limit $\pi/2g^2_5$ for any finite $k$.

\begin{figure}[thb]
  \centering
	\subfigure[]{\label{fig:rho400}
	\includegraphics[width=.45\textwidth]{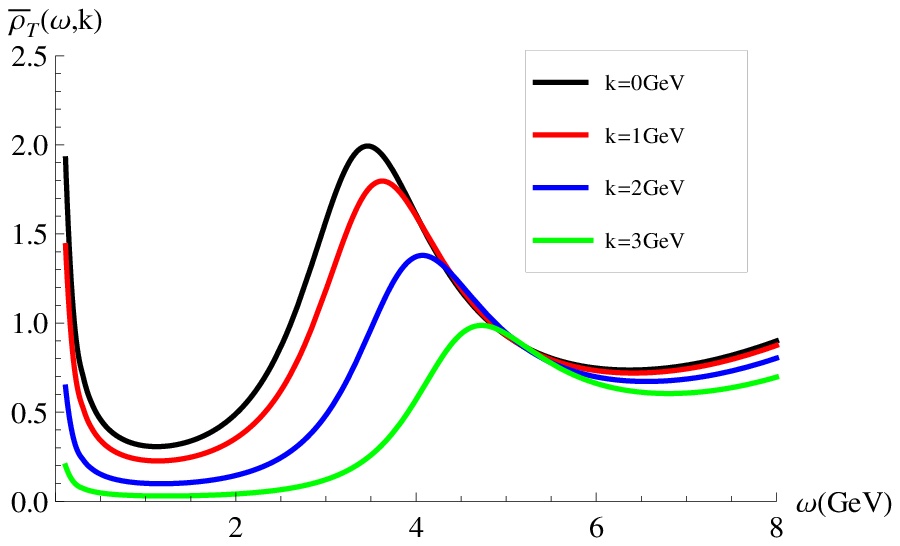}}
	\subfigure[]{\label{fig:rhoTvsL400}
	\includegraphics[width=.45\textwidth]{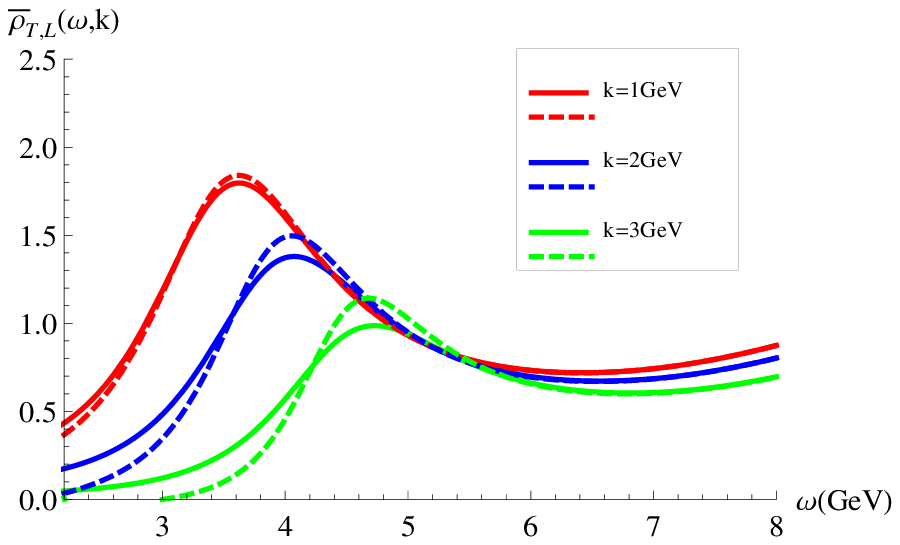}}
\caption{
  \label{fig:rhoT}
Left: Rescaled spectral functions in transverse channel,
$\bar{\rho}_{T}(\o, k)$, defined by \eq\eqref{eq:rho-bar} at $T=0.74\, T_{\text{D}}(k=0)$ where $T_{\text{D}}(k=0)$ is the dissociation temperature at zero momentum.
Black, red, blue and green curves are corresponding to $k =0,1,2,3$ GeV, respectively.
%Dashed curve shows the change of J/$\psi$ peak with changing momentum.
Right: Comparison between rescaled spectral functions, $\bar{\rho}(\o, k)$, in transverse channel (solid) and longitudinal channel (dashed) at $T=0.74\, T_{\text{D}}(k=0)$.
Red, blue and green curves correspond to momenta $k =1,2,3$ GeV in each channel.
}
\end{figure}

Figure \ref{fig:rho400} plots the rescaled spectral functions $\bar{\rho}_{T}(\o, k)$ from the ``shift and dip" model in the transverse channel at $T= 0.74\, T_{\text{D}}(k=0)$ where $T_{\text{D}}(k=0)$ is the J/$\psi$ dissociation temperature at zero momentum (cf.~Sec.~\ref{sec:dissociationT} below or Ref.~\cite{Grigoryan:2010pj} for the discussions on determination of $T_{\text{D}}(k=0)$).
We first observe  that the
$\psi'$ has already merged into the continuum even at zero momentum. Thus further discussions on $\psi'$
at finite momentum will not be made.
Secondly, the peak location corresponding to the J/$\psi$ metastable state moves to larger $\omega$ while the peak height is attenuated as momentum increases.
The sharp peak at $\omega = 0$ in the transverse channel is associated with the conductivity of
the current of charm charge when $k = 0$.
At finite $k$, this peak, i.e., $\rho(\o, k)/\o$, decreases with increasing momentum.

To compare the transverse and longitudinal spectral functions, we plot $\bar{\rho}_T(\o, k)$ and $\bar{\rho}_L(\o , k )$ in Fig.~\ref{fig:rhoTvsL400} at $T= 0.74\, T_{\text{D}}(k=0)$ for $k = 1 , 2 , 3$ GeV.
For frequencies around and larger than the peak region,
the transverse and longitudinal spectral functions  are quite similar.
In the peak region, the shift in mass with momentum is also similar between the two modes.  The
height of the peak is less attenuated while it is narrower in the longitudinal mode compared with the transverse mode.

\subsection{Dispersion relation and spectral width \label{sec:dispersion}}
To quantify the properties of the spectral function, we can calculate the poles of the retarded Green's function,
denoted by $\Omega_{T,L}(k,T)$.
On the field theory side, the real part of the pole is attributed to the peak mass while the negative of the imaginary
part with the width.
In the literature of holographic correspondence,
those poles are referred to as quasi-normal modes \cite{Nunez:2003eq,*Kovtun:2005ev}
which can be determined by solving
$V(\epsilon, \O, k ) = 0$, using
the solutions of \eq\eqref{eq:V} and boundary conditions given in \eq\eqref{eq:bc-V} for any given $k$.
At zero temperature and zero momentum,
$\O_{T}(k=0,T=0)=\O_{L}(k=0,T=0) = M_{J/\psi}$.
We will present numerically determined $\O_{T, L}(k, T)$ below.

\begin{figure}[htb]
  \centering
	\subfigure[]{\label{fig:Tdispersion}
	\includegraphics[width=.45\textwidth]{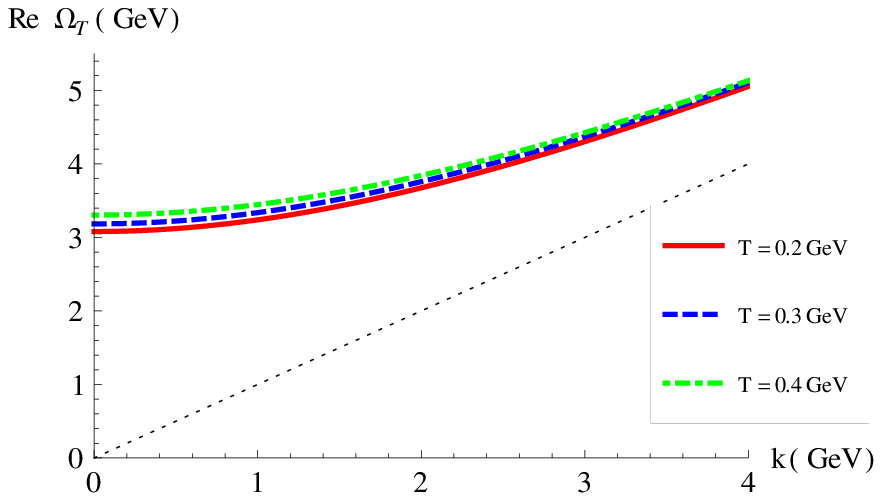}}
	\subfigure[]{\label{fig:Ldispersion}
	\includegraphics[width=.45\textwidth]{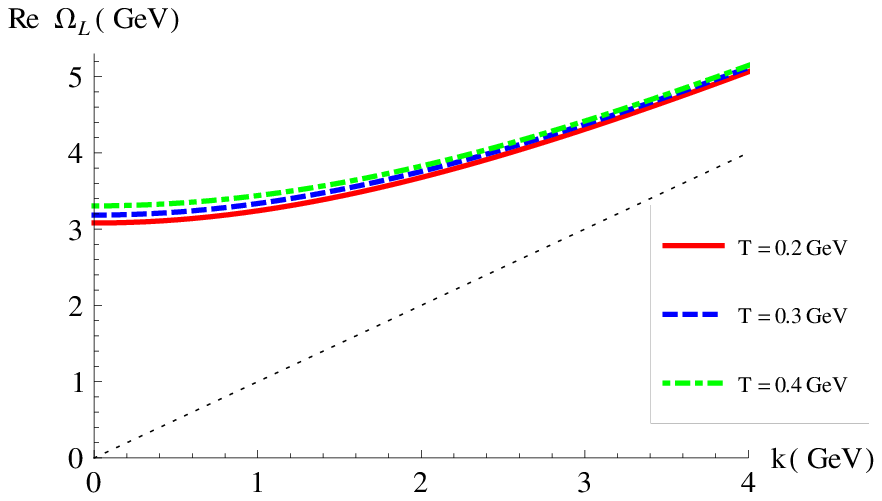}}
\caption{
\label{fig:dispersion}
Left: Dispersion relation for J/$\psi$ in the transverse channel.
Right: Dispersion relation for J/$\psi$ in longitudinal channel.
Red,blue and green curves are corresponding to $T=200, 300, 400$ MeV respectively.
The dotted curve is corresponding to the light cone.
}
\end{figure}

Figure \ref{fig:dispersion} plots the pole mass of J/$\psi$,
$\text{Re} \,\O_{T, L}(k, T)$, as a function of momentum.
These curves represent the in-medium dispersion relation of J/$\psi$ calculated from the ``shift and dip" model.
They are
plotted at three different temperatures, from bottom to top, $T = 200$ MeV
(red), $T =300$ MeV (blue), and $T = 400$ MeV (green).
The dispersion relation shows little temperature dependence and no indication of a subluminal limiting velocity.
Moreover, the transverse and longitudinal modes reveal nearly identical dispersion relations.
This is consistent with our results on spectral densities presented in last section(cf.~Fig.~\ref{fig:rhoTvsL400}).
Numerically, the relative difference between $\O_{T}(k,T)$ and $\O_{L}(k,T)$ at fixed $k, T$
is within a few percent for all temperatures above $T_c$ and momenta which we are considering.

\begin{figure}[htb]
  \centering
	\subfigure[]{\label{fig:width-holo}
	\includegraphics[width=.45\textwidth]{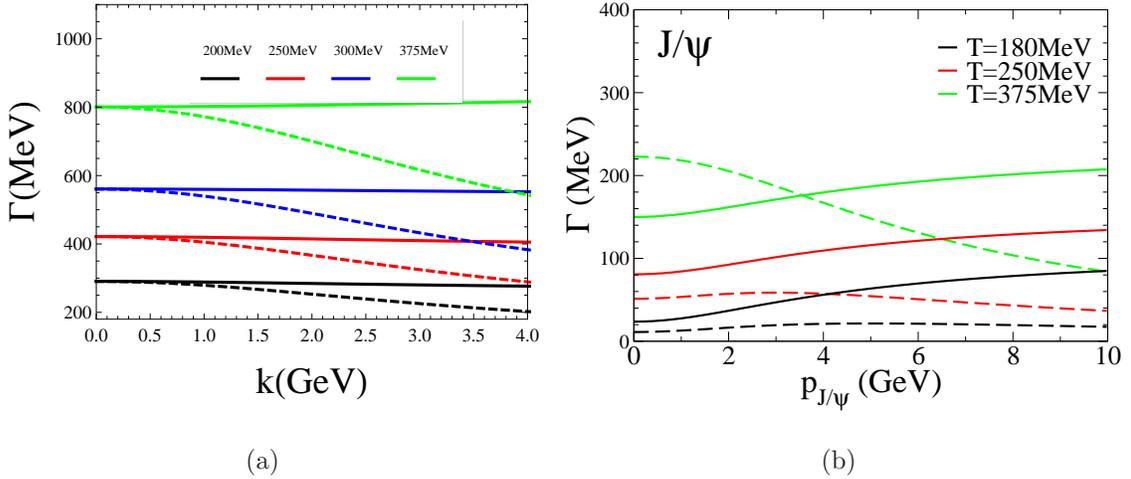}}
	\subfigure[]{\label{fig:width-perp}
	\includegraphics[width=.45\textwidth]{width-perp.eps}}
	\caption{
	\label{fig:width}
Left: Momentum dependence of the total width of J/$\psi$ in transverse channel.
Black,red, blue and green curves are corresponding to $T=200, 250, 300, 375$ MeV respectively.
Dashed curves are corresponding to those from the longitudinal channel.
Right: Momentum dependence of quasi-free (solid line)
and gluo-dissociation rates (dashed line) for J/$\psi$ calculated from perturbative QCD at different temperatures from Ref.~\cite{Zhao:2007hh} (see also Ref.~\cite{Rapp:2008tf}).
	}
\end{figure}

The quasinormal modes can also be used to determine the peak width of J/$\psi$:
$\Gamma_{L , T} = - \text{Im}\O _{L, T}$.
Figure \ref{fig:width-holo} presents the width $\Gamma_{L, T}(k, T)$ as a function of momentum for four different temperatures for both transverse and longitudinal channel.
In the transverse channel,
the width is a slowly varying function of momentum for all temperatures we are considering.
On the other hand,
the width will decrease as momentum increases in the longitudinal channel.
This numerical behavior is consistent with the qualitative observations from the last section.

\subsection{
\label{sec:dissociationT}
Momentum dependence of the dissociation temperature }
We now study the momentum dependence of the dissociation temperature.
In order to determine the dissociation temperature, $T_D$, of J/$\psi$, one must select a criteria.
In general, $T_D$ is associated with the temperature when the J/$\psi$ peak is no longer present
in the spectral function. Yet a more  specific criteria is useful.
We will
follow Ref.~\cite{Grigoryan:2010pj} and
define the dissociation temperature $T_D(k)$ at momentum $k$ as the temperature below which the height of the J/$\psi$ peak, $H_{J/\psi}$, of the rescaled spectral functions
$\bar{\rho}(\o, k)$
is smaller than some threshold value, say $H_D$.
Of course, there is no unique choice of $H_D$,
similar to the situation that there is no precise definition of screening length below which J/$\psi$ will dissociate.
In Ref.~\cite{Grigoryan:2010pj},
the simplest criterion
\begin{equation}
\label{eq:HD}
H_D = 1
\end{equation}
is used to quantify the dissociation temperature.
The idea behind this criterion is that
a J/$\psi$ state can be considered as being merged into continuum
if its peak is lower than the asymptotic value of the rescaled spectral function
$\bar{\rho}(\o, k)$, and thus be considered dissociated.
We will follow the criterion of \eq\eqref{eq:HD} here.
It should be also pointed out that
the peak height calculated in the present holographic model
is a monotonic function of both momentum and temperature,
which makes it suitable as a criterion candidate.
At zero momentum, the rescaled spectral density calculated from ``shift and dip" model $\bar{\rho}(\o, k = 0)$
supports a J/$\psi$ peak up to at least $500$ MeV \cite{Grigoryan:2010pj}.
Using the criterion in \eq\eqref{eq:HD}, the
zero momentum dissociation temperature is found to be
\begin{equation}
T_\text{D}(k=0) = 540\, \text{MeV}\, .
\end{equation}
This is consistent with lattice measurements reported in Refs.~\cite{Jakovac:2006sf,Aarts:2007pk} that J/$\psi$ will survive up to $2 \sim 2.4\, T_c$ with $T_c=190$ MeV but is higher than the lattice results of Ref.~\cite{Ding:2012sp} that J/$\psi$ are found to dissociate already at $1.5\, T_c$.

\begin{figure}[htb]
  	\subfigure[]{\label{fig:Tdq}
	\includegraphics[width=.45\textwidth]{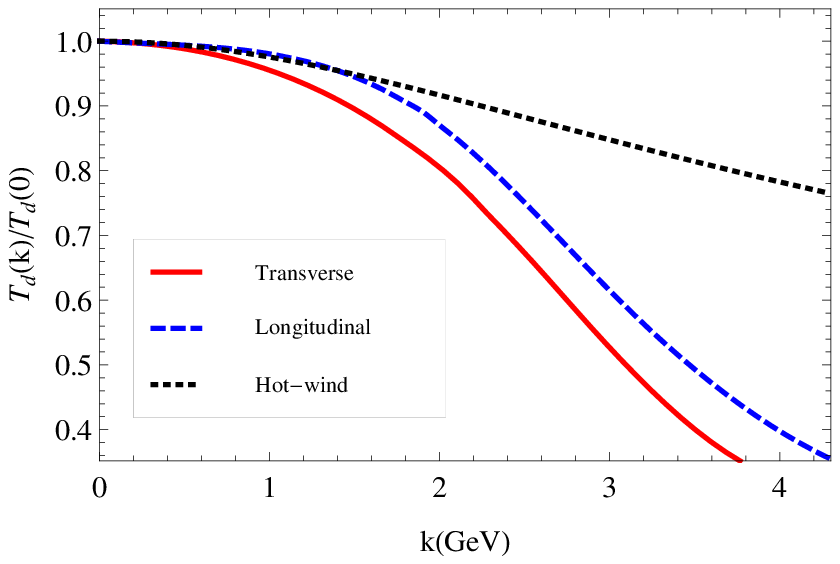}}
	\subfigure[]{\label{fig:qdT}
	\includegraphics[width=.45\textwidth]{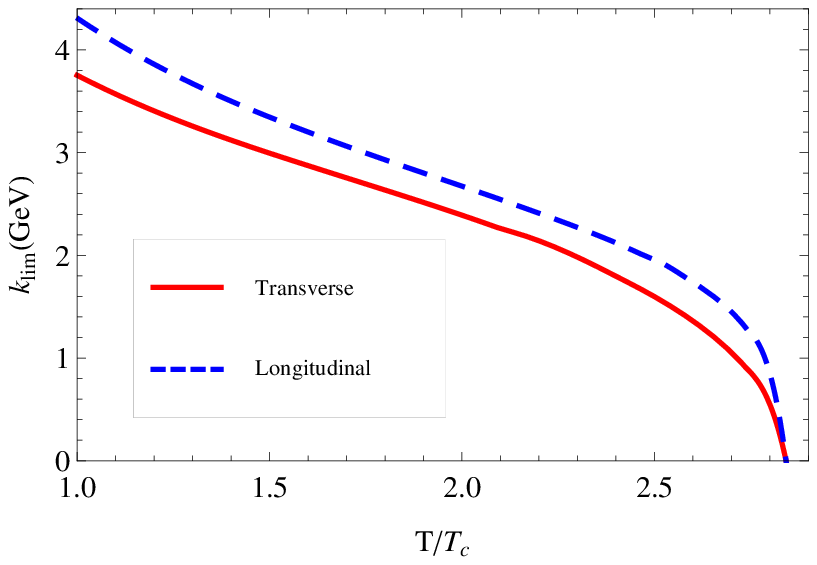}}
\caption{
  \label{fig:T-q}
  Left panel:
Momentum dependence of the dissociation temperature $T_\text{D}(k)/T_\text{D}(k=0)$
in the range $T_\text{D}(k=0)\geq T_\text{D}(k) \geq T_\text{c}$ in the transverse channel (red, solid) and  the longitudinal channel (blue, dashed) with $T_c = 190$ MeV.
The dotted curve shows the momentum dependence of the dissociation temperature
based on the holographic analysis of color screening length \cite{Liu:2006nn} assuming the approximate scaling in \eq\eqref{eq:hotwindscaling} and a relativistic on-shell relation.
Right panel:
The limiting momentum $k_{\text{lim}}$ as a function of temperature $T$ in the transverse (red) and  the longitudinal (blue) channels with $T_c = 190$ MeV.
}
\end{figure}

Using this selection process, $T_D$ can be determined as
as a function of momentum. This is plotted in Fig.~\ref{fig:Tdq}.
Observe that, as expected,
as momentum increases, the dissociation temperature decreases.
Furthermore, the transverse mode decreases faster than the longitudinal mode with
momentum. This is in accordance to the difference in the peak height of the spectral functions
seen above.
At a finite momenta, the dissociation temperature becomes
equal to $T_c$. Though there is no manifestly inherent transition in this model
at this momentum, on physical grounds $T_D$ should be greater than $T_c$. One can also ask the question,
for a given temperature, what is the largest momentum J/$\psi$ could have before it dissociates. This limiting
momentum, $k_{\rm lim}$, is shown in Fig.~\ref{fig:qdT}. The maximum value $k_{\rm lim}$ can achieve is at $T=T_c$, {\it i.e.} when
$T_D(k_{\rm lim}^{\rm max} ) = T_c$. This condition
suggests that the momentum of charmonia in-medium could be limited to a maximum
value unrelated to any dispersion relation. For $T_c \sim 190$ MeV, we find the maximum momentum
to be $k_{\rm lim}^{\rm max} = 3.8$ GeV for the transverse mode and $k_{\rm lim}^{\rm max}=4.2$ GeV for the longitudinal mode.
The implications of this momentum will be discussed in subsequent sections.

\section{Discussion \label{sec:discussion}}
In this section, we will interpret and discuss the results which were represented in the previous sections,
while also placing them in the context of existing literature.

\subsection{J/$\psi$ peak height}

\begin{figure}[htb]
  \centering
	\subfigure[]{\label{fig:residue-holo}
	\includegraphics[width=.45\textwidth]{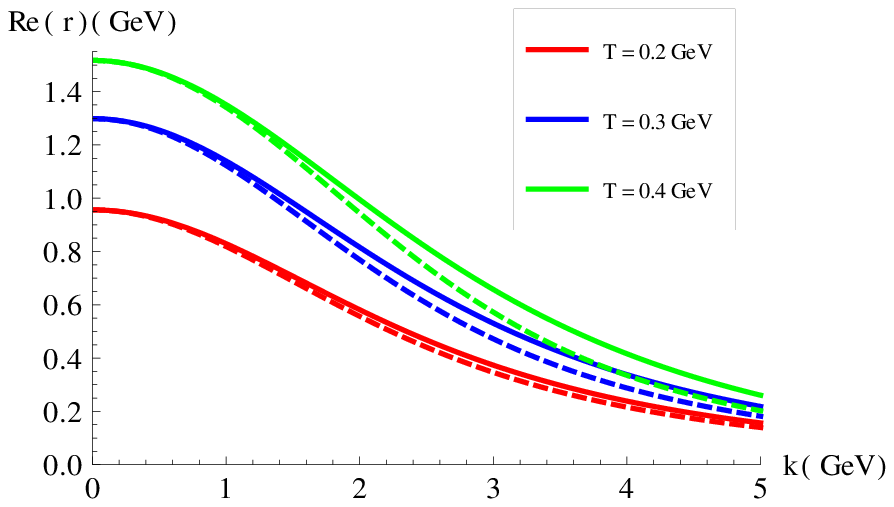}}
	\subfigure[]{\label{fig:height-holo}
	\includegraphics[width=.45\textwidth]{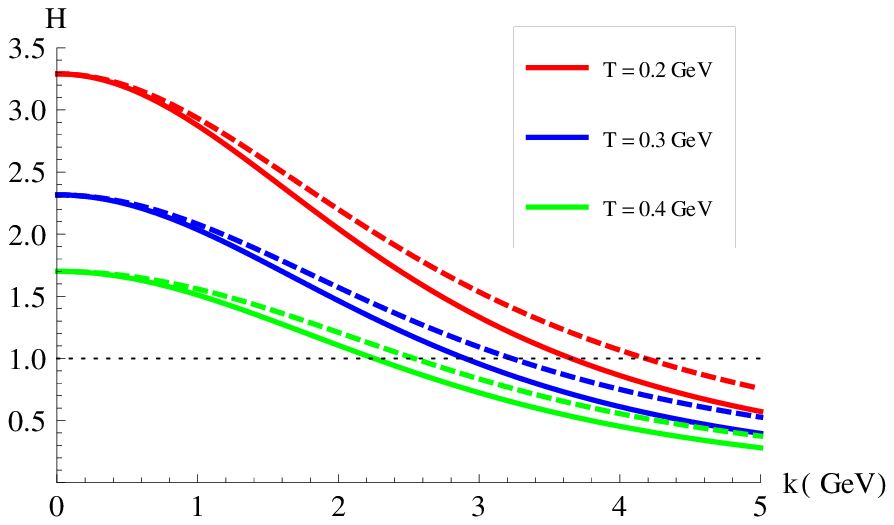}}
	\caption{
	\label{fig:height-residue}
Left: Momentum dependence of the real part of the residue of J/$\psi$ in transverse channel.
Red,blue and green curves are corresponding to $T=200, 300, 400$ MeV respectively.
Dashed curves are corresponding to those from the longitudinal channel.
Right: Momentum dependence of the height of J/$\psi$ in transverse channel.
Red, blue and green curves are corresponding to $T=200, 300, 400$ MeV respectively.
Dashed curves are corresponding to those from the longitudinal channel.
	}
\end{figure}

The most important result presented in this paper is the decrease in dissociation temperature with momentum.
This momentum dependence is due to the decrease of the height of the J/$\psi$ peak in both the transverse
and longitudinal modes. This behavior is qualitativley in agreement with results from lattice \cite{Ding:2012sp}, holography \cite{Liu:2006he,Mateos:2006nu,Ejaz:2007hg} and heavy quark effective theory \cite{Escobedo:2011ie}.
Moreover, the height of the J/$\psi$ peak is related to the width $\Gamma$ and the (rescaled) residue associated with J/$\psi$ pole of $G^{\text{R}}_{\mu\nu}(\o, \vk)$, i.e. ,
\be
H\equiv
\frac{\text{Re}\,(r_{J/\psi})}{\Gamma}\, .
\ee
Here the rescaled residue $r_{J/\psi}$ corresponding to the J/$\psi$ pole is defined by
\be
r_{J/\psi} \equiv
\frac{2g^2_5}{\pi\,\Omega^2_{J/\psi}}\lim_{\o\to \Omega_{J/\psi}}\(\o -\Omega_{J/\psi}(k)\)\Pi^{\text{T,L}}(\o, k)\, .
\ee
It is interesting to identify whether the momentum dependence of the height is associated with the width or the residue.
In holographic models, the width $\Gamma$ and the rescaled residue $r_{J/\psi}$ can be easily calculated.
The momentum dependence of $\Gamma$ and $r_{J/\psi}$ are ploted in Figs.~\ref{fig:width-holo} and~\ref{fig:residue-holo} respectively.
Comparing those figures with the momentum dependence of J/$\psi$ peak height $H$ in Fig.~\ref{fig:height-holo},
we found that in the present model,
the momentum dependence of J/$\psi$ peak height $H$ is dominated by the momentum dependence of the rescaled residue $r_{J/\psi}$.
Indeed, in transverse channel, the width is a slowly varying function of $k$ while in the longitudinal channel,
the width even decreases with growing momentum.
However, due to relatively rapid decreasing of the rescaled residue $r_{J/\psi}$,
the height $H$ is attenuated in both channels. Therefore we must conclude that the dissociation temperature
is more sensitive to the residue rather than the width of the peak.

\subsection{Total width}
\begin{figure}[htb]
 \centering
	\includegraphics[width=.45\textwidth]{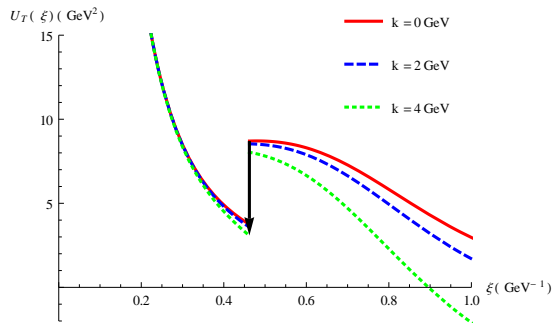}	
\caption
{
\label{fig:potential}
Holographic potential \eq\eqref{eq:Schroedinger-T} at $T=400$ GeV with $k = 0$ GeV (red,solid),
$k = 2$ GeV (blue,dashed), $k =4$ GeV (green, dotted) respectively.
The black arrow indicates the negative delta-function in ``holographic potential" at $z = z_d$.
}
\end{figure}

One striking result was the different behaviors of the J/$\psi$ width between the transverse
and longitudinal channels.
One can understand the behavior of J/$\psi$ width in the transverse channel shown in Fig.~\ref{fig:width-holo}  from the gravity
side of the duality.
The finite temperature and momentum state corresponding to charmonium is represented by
a metastable quantum state which satisfies an equation of Schr\"{o}dinger form, Eq.~(\ref{eq:Schroedinger-T}),
with the holographic potential \eq\eqref{eq:U-B-T}.
Therefore,
one can apply one's normal intuition concerning quantum mechanical systems.
In Fig.~\ref{fig:potential},
the holographic Schr\"{o}dinger potential is plotted at $T=400$ MeV and a variety of
momenta.
One can clearly see that the potential barrier at higher values of $\xi$,
the ``Sch\"{o}dinger coordinate'' defined in \eq\eqref{eq:Liouville-T},
decreases in
magnitude and narrows with momentum.
Therefore, at higher momentum,
it is easier for the quantum state to tunnel through this barrier and thus decay.
Consequently,
the absolute value of the imaginary part of $(\O_{J/\psi})^2$,
$ 2 (Re\O_{J/\psi}) \Gamma$,
will increase with higher momentum.
On the other hand,
as we see from Fig.~\ref{fig:Tdispersion},
$Re\O_{J/\psi}$ is also growing with increasing momentum.
As a result,
the change of the total width, $\Gamma$,
in transverse channel is relatively slow. Unfortunately,
a similar argument for the longitudinal mode is not as clean because both the Schr\"{o}dinger
potential and its coordinate are complex.
However,
the different behaviors of the J/$\psi$ width may be understood as a ``Doppler-like" effect analogous to that seen in the weakly coupled calculation of Ref.~\citep{Escobedo:2011ie}.
The difference between the two channels may also imply
a coupling between the spin and the gluon fields on the field theory side.

\begin{comment}
Unfortunately,
a similar argument for the longitudinal mode is not as clean because both the Schr\"{o}dinger
potential and its coordinate are complex.
However, the behavior of the total width in the longitudinal channel as shown in Fig.~\ref{fig:width-holo} is not unexpected.
Remember charmonium is a $\bar{c}c$ pair separated by a fixed distance.
Therefore,
in terms of the ``charm" charge, J/$\psi$ acts as a dipole oriented along the axis which connects the
pair of quarks.
For the transverse mode, this dipole (and thus the spatial extent of the particle)
is perpendicular to the direction of motion.
However,
for the longitudinal mode,
the dipole is
directed parallel to its motion.
Thereby, the spatial extent of the particle is Lorentz contracted
relative to the transverse mode.
This contraction makes the $\bar{c}c$ pair become closer together
and increases the binding of the state.
Thus with increased momentum, the longitudinally polarized J/$\psi$
becomes more stable resulting in a decrease in the spectral peak width.
Hence the difference in the two channels.
\end{comment}

The numerical results for the peak width as a function of momentum can also be compared with the expectations from QCD. In the QGP, J/$\psi$ may be (inelastically)
dissociated via either interactions with gluons, i.e. gluon dissociation reactions,
$J/\psi + g \rightarrow c + \bar{c}$, or collisions with partons in the medium, i.e.
quasi-free dissociation, $X + J/\psi \rightarrow X + c + \bar{c} (X=g, q, \bar{q})$ \cite{Rapp:2008tf,Zhao:2007hh}.
These two mechanisms have
different momentum dependence.
Due to increasing color screening length,
gluon dissociation
tends to be a decreasing function of momentum.
On the other hand,
the quasi-free dissociation rate always increases with momentum due to a smoothly increasing cross section with increasing center-of-mass energy.
The momentum dependence of each of these process has been previously calculated in Ref.~\cite{Zhao:2007hh}
 and illustrated in Fig.~\ref{fig:width-perp}.
The combination of the two process results in a width which increases with momentum
at lower temperatures and decreases with momentum at higher temperatures.
If we average over the polarization states in the holographic model, we observe a
qualitative similar trend, that as one increases the temperature, the width decreases
more and more with momentum because of the behavior of the longitudinal channel. The absolute size of the
width for the current model is larger than the QCD predictions, but this may be because
the holographic width includes both inelastic and elastic scattering whereas the QCD width
only includes the former. Nevertheless, it is interesting to note the qualitative similar momentum behavior
between strongly and weakly coupled calculation.

\subsection{Lattice}

\begin{figure}[htb]
  \centering
	\subfigure[]{\label{fig:srho400}
	\includegraphics[width=.31\textwidth]{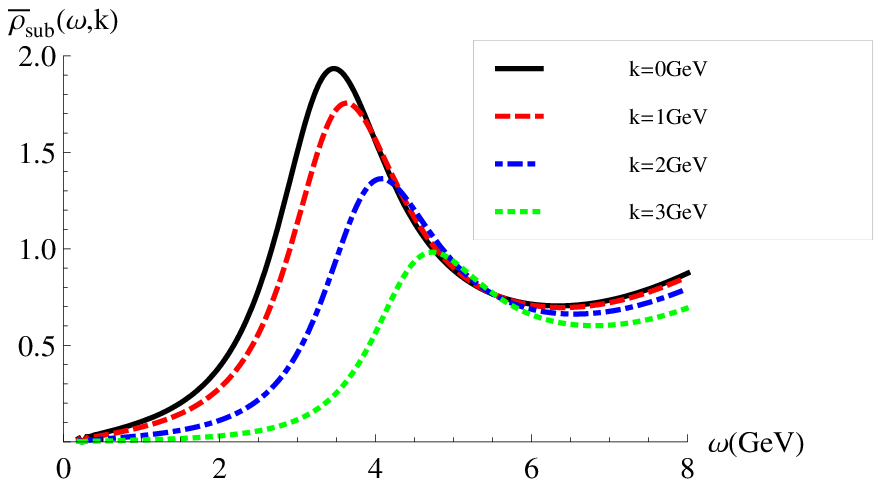}}
	\subfigure[]{\label{fig:srho500}
	\includegraphics[width=.31\textwidth]{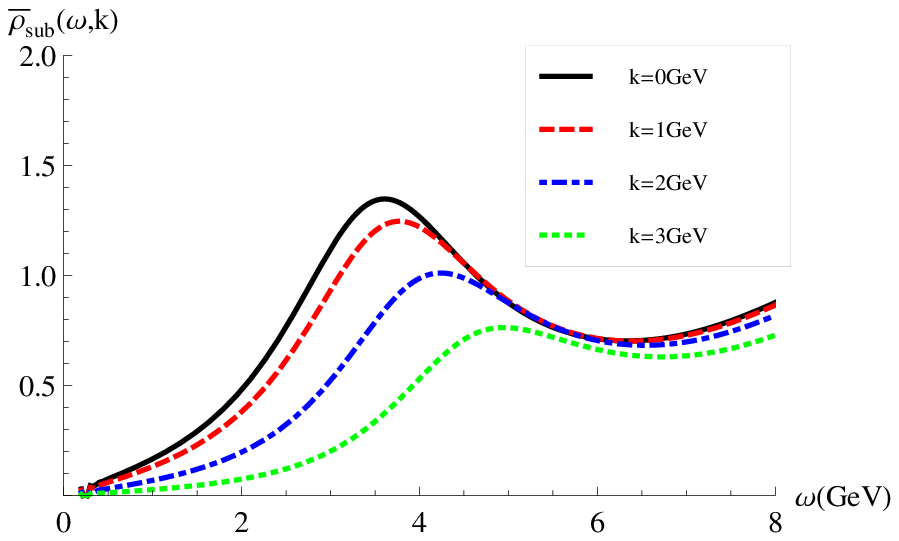}}
	\subfigure[]{\label{fig:srho600}
    \includegraphics[width=.31\textwidth]{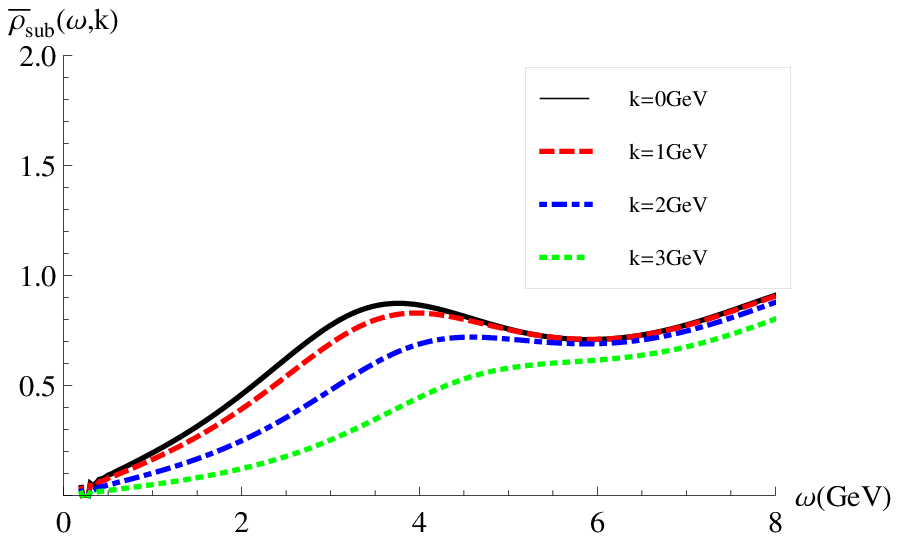}}
\caption{
  \label{fig:rhosub}
Momentum dependence of $\bar{\rho}_{\text{sub}}(\o, k)$
defined by \eq\eqref{eq:rho-bar-p} at $T=0.74\, T_{\text{D}}(k=0)$(left), \,$T=0.94\, T_{\text{D}}(k=0)$(middle) and $T= 1.1\, T_{\text{D}}(k=0)$(right).
Black, red, blue and green curves are corresponding to $k =0, 1, 2, 3$ GeV respectively.
}
\end{figure}

\begin{figure}[htb]
  \centering
	\includegraphics[width=.45\textwidth]{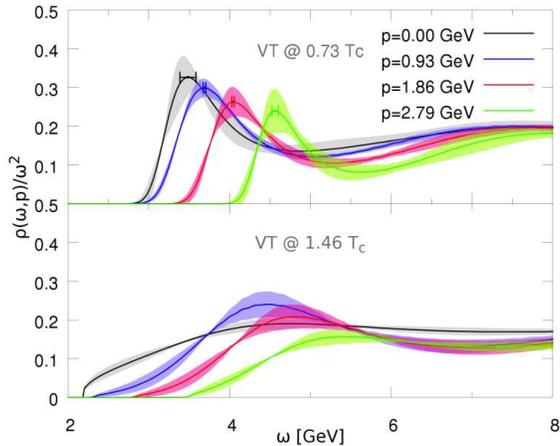}
\caption{
	\label{fig:latticerho}
Lattice measurements of momentum-dependent (rescaled) spectral functions in transverse channel in Ref.~\cite{Ding:2012pt} .
}
\end{figure}

One motivation of the present paper is to provide complementary information for the lattice measurement of J/$\psi$ spectral function.
One may note that as the width of the transport peak is inversely proportional to the mean free path,
spectral functions calculated from holography generically will have a relatively broadened hydrodynamic peak.
To minimize the effects of such broadened transport peaks on the J/$\psi$ peaks we are interested in
and to make a closer comparison with lattice results, we define a subtracted rescaled spectral function,
\begin{equation}
\label{eq:rho-bar-p}
\bar{\rho}_{\text{sub}} (\o, k)
= \bar{\rho}(\o, k) - \frac{1}{\o}
\lim_{\omega\to0} \[\rho(\omega, k)/\omega\]\, .
\end{equation}
At zero momentum,
$\bar{\rho}_\text{sub}(\o, 0)$ is just the rescaled spectral function with the subtraction of the transport peak\footnote{The definition of $\bar{\rho}_{\text{sub}}(\o, k)$ in \eq\eqref{eq:rho-bar-p} is motivated by a sum-rule type model-independent relation which holds for a large class of field theories with a generic gravity dual description
that spectral functions can be represented as a \textit{convergent} sum
over simple poles of the corresponding retarded Green's function in terms of conductivity $\sigma$ and the residues \cite{Stephanov:2011wf}. }.
We plot $\bar{\rho}_{\text{sub}}(\o, k)$
in the transverse channel for three representative temperatures $T = 0.74, 0.94, 1.1\,T_{\text{D}}(k=0)$ in Fig.~\ref{fig:rhosub}.
As we have mentioned earlier in Sec.~\ref{sec:dissociationT},
there is a discrepancy in zero momentum dissociation temperature $T_{\text{D}}(k=0)$ between the holographic
 calculations and the lattice literature.
To minimize the ambiguities due to the difference in $T_{\text{D}}(k=0)$ and to highlight the momentum dependency,
results presented in Fig.~\ref{fig:rhosub} should be compared with existing and future lattice measurements at the same temperature relative to the zero momentum dissociation temperature, i.e., $T/T_{\text{D}}(k=0)$.
Specifically, the transverse spectral density measured at $T=1.46 T_c$ on the lattice by Ref.~\cite{Ding:2012pt},see Fig.~\ref{fig:latticerho},
should be compared with the holographic calculation at $T=1.1 T_D(k=0)$, i.e., Fig.~\ref{fig:srho600}.
Furthermore, the different behavior of the transverse and longitudinal channels observed here is qualitatively confirmed by the lattice calculations of   Ref.~\cite{Oktay:2010tf}.

Indeed,
the momentum dependence of spectral density from the ``shift and dip" model at that relative temperature is qualitatively consistent with the lattice reconstructed spectral function reported recently in Ref.~\cite{Ding:2012pt}(see also Fig.~\ref{fig:latticerho}).
We take this agreement as one of the evidences
that the ``shift and dip" model is consistent with the in-medium properties of charmonium.

\subsection{Limiting velocity}

The dispersion relation results indicate little temperature dependence and no indication of a
subluminal limiting velocity, i.e., $\omega \sim v_g k$ with $v_g=1$ for large $k$.
This is consistent with the analysis of
Ref.~\cite{Festuccia:2008zx} which used a WKB approximation to determine the dispersion relation for large $k$ in a class of
holographic models.
However, it is contrary to the work of \cite{Ejaz:2007hg} which was based on
a ``top-down" perspective, and treated charmonium as an excitation of a D-branes.
There they found that although for large $k$,
$\o \sim v_g k$,
$v_g$ depends on the temperature and is smaller than $1$.
The observation of a subluminal limiting velocity in ``top-down" models was also established in Ref.~\cite{Mateos:2006nu}.
It was suggested that the presence of this subluminal limiting velocity was associated with charmonium melting, i.e.,
if the group velocity of a J/$\psi$ were greater than that subluminal group velocity,
it would dissociate.
Moreover, as noticed in Ref.~\cite{CasalderreySolana:2008ne},
if dispersion relation curve of the heavy quarkonia crosses the light-cone at some point in the frequency space,
photon production rate per energy will have a peak accordingly.
Based on the above observations
and generic features of dispersion relation of quarkonia found in ``top-down" holographic approach \cite{Mateos:2006nu, Ejaz:2007hg},
authors of Ref.~\cite{CasalderreySolana:2008ne}
predicted
a photon production peak between $3$ and $5$ GeV due to J/$\psi$ in heavy ion collisions.

We can understand the discrepancy between these other models and the results presented here
in the following way. In each of the ``top-down" constructions
in Refs.~\cite{Mateos:2006nu,Ejaz:2007hg},
D-branes lie completely outside the horizon  in a ``Minkowski embedding.''
In the large $k$ limit, the wave function of a meson is localized at the tip of the branes.
Consequently, the limiting group velocity of this meson approaches the local speed of light at the tip of the
branes and is subluminal compared with the speed of light in flat space.
For the class of models
including the ``shift and dip" model,
the bulk geometry extends to the horizon.
As conjectured in Ref.~\cite{Myers:2008cj} and shown in Ref.~\cite{Festuccia:2008zx} via WKB analysis,
in such geometries,
the support of quasinormal modes becomes increasingly
focused at the asymptotic AdS boundary,
where the local speed of light is asymptotically $1$.
    Thus the limiting group velocity is $1$ as well.
Furthermore,
because the dispersion relation of the current model does not cross the light cone,
 we would not expect to observe a peak in the photon production due to J/$\psi$.
This is then consistent with the recent experimental report \cite{Adare:2008ab}
on photon production with Au-Au collision at $\sqrt{s_{NN}}=200$ GeV by the PHENIX collaboration.
 This suggests that in the future holographic constructions,
certain constraints from experiment on limiting velocity should
be taken into account.

%\subsection{Width from perturbative calculations}

\subsection{Hot wind}
Next we come to the dissociation temperature as a function of momentum.
In the pioneering work of Ref.~\cite{Liu:2006nn},
the Wilson loop, thus the heavy quark effective potential,
and the color screening length in a ``hot wind" have been calculated using holographic techniques.
The authors of Ref.~\cite{Liu:2006nn} found that the dissociation temperature $T_D$,
defined as the temperature such that the screening length is the same order as the size of the quarkonia,
decreases as the relative velocity $v_r$ between the heavy quarkonia and the medium increases.
To good accuracy,
the dissociation temperature scales with relative velocity $v_r$ as:
\begin{equation}
\label{eq:hotwindscaling}
T_D(v_r) \approx (1 - v_r^2)^{1/4} T_D(v_r = 0)\, .
\end{equation}
The scaling behavior of \eq\eqref{eq:hotwindscaling} may be understood
from relativistic kinematics \cite{CasalderreySolana:2011us}.
Considering a meson moving through the plasma with velocity $v_r$,
it experiences a higher energy density, boosted by a factor of $\gamma^2= (1-v_r^2)^{-1}$.
As energy density is proportional to $T^{4}$, the effective temperature is boosted by a factor of $\sqrt{\gamma} = (1 - v_r^2)^{-1/4}$. This parallels the previous observation that the limiting velocity of
Refs.~\cite{Mateos:2006nu, Ejaz:2007hg} had the
same dependence on temperature as the dissociation temperature for the ``hot-wind" model had on momentum.

For the sake of the comparison,
we convert \eq\eqref{eq:hotwindscaling} into the momentum dependence of $T_D(k)$ assuming relativistic kinematic relation $\o(k) = \sqrt{k^2 + M^2_{J/\psi}}$
and plot $T_D(k)$ verses $k$ in Fig.~\ref{fig:Tdq} (dotted line, see also Fig.~3 of \Ref\cite{Liu:2006nn}).
Qualitatively, we see from Fig.~\ref{fig:Tdq} that the dissociation temperature $T_D$ determined
either from the screening length or the height of J/$\psi$ peak
is a decreasing function of momentum with the present approach decreasing faster than the ``hot wind"
analysis.
This difference may be attributed to the fact that
in addition to color screening,
other dissociation mechanisms such as thermal broadening,
are also incorporated in the current study.
The present analysis on dissociation temperature provides new insights,
which are complementary to previous ones,
into the momentum dependence of J/$\psi$ dissociation.

\subsection{
\label{sec:RAA}
Nuclear modification factor, $R_{AA}$, and a maximum momentum}
Throughout this paper,
we have calculated the attenuation of J/$\psi$ peak height with increasing momentum
and determined the dissociation temperature $T_{\text{D}}(k)$ as a function of J/$\psi$ momentum $k$ based upon height
of the spectral peak.
This was a convenient choice because it was a monotonic function of both
temperature and momentum,
and thereby removed ambiguities of selecting the dissociation temperature
besides that which was associated with the criterion.
Yet the relevant experimental observable for J/$\psi$ suppression is the nuclear modification factor,  $R_{AA}$.
It is not  clear {\it a priori} which aspect of the equilibrium properties of charmonium is most important
to determine $R_{AA}$. For this discussion, we will focus on the dissociation temperature.
If $T_D$ were the dominated factor in determining $R_{AA}$,
we would speculate, from the results presented here, that $R_{AA}$ should decrease as a function of momentum. Yet, as we have indicated above, this decrease should not continue to arbitrarily small temperatures; the dissociation temperature needs to be restricted to $T_D > T_c$. We have shown in Sec.~\ref{sec:dissociationT} that this condition implies that there is a maximum momentum which J/$\psi$ could achieve before dissociation, namely $k_{\rm lim}^{\rm max}$. If J/$\psi$ is completely dissolved above this momentum, this would constitute a maximum level of suppression. Therefore, $R_{AA}$ should decrease with increasing momentum
until it approaches this maximum level of suppression at some critical momentum set by $k_{\text{lim}}^{\rm max}$,
and remain relatively constant after that.
More precisely,
that implies that (a) $R_{AA}$ would decrease with increasing rapidity
and
(b) the $p_t$ that a saturated suppression begins would also decrease with increasing rapidity.
This is exactly what is observed at RHIC \cite{Adare:2006ns,Adare:2008sh}.

We would also like to remind the reader that though
we have emphasized throughout the differences,
if any,
between the transverse and longitudinal modes, this distinction is not
experimentally possible at this time.
Thus for any comparison with experiment one would
need to average over the polarization states. The holographic model described here suggests  $k_{\rm lim}^{\rm max} \sim 3-4$ GeV. Remember that this is given in the rest frame of the plasma,
so in the lab frame,
this would be different depending on the velocity of the expanding QGP. Additionally, in the present paper,
we have been considering charmonia which are in equilibrium with the medium.
This is not the only source of charmonia yields observed in experiment.
Therefore our results would only apply to $R_{AA}$ in the $p_t$ window where the yields are dominated by those J/$\psi$
in equilibrium with the QGP. Lastly, we have described the transition from un-dissociated to dissociated as being rather sharp while physically this is not the case. This would simply smear out the effect and the onset of a limiting momentum.

The momentum dependence of $R_{AA}$ speculated here is based solely on the momentum dependence of $T_D$.
Thus the existence of a limiting momentum is not a unique
feature of ``shift and dip'' model.
Any model whose dissociation temperature decreases monotonically
with momentum would see a similar feature,
including  other holographic models and pQCD calculations.
The difference,
from the phenomenological point of view,
would be the numerical value of that limiting momentum, $k_{\text{lim}}^{\rm max}$.
Remarkably,
the limiting momentum obtained from the present model is much lower than previous holographic models \cite{Liu:2006he,Ejaz:2007hg}.
What is more,
if we take $k_{\text{lim}}^{\rm max}$ as a crude estimate of the $p_t$ for charmonia at midrapidity then the results obtained here are quantitatively in agreement with experiment results \cite{Adare:2006ns,Adamczyk:2012ey}
that a saturated suppression begins around $p_t = 3\sim 4$GeV.

\section{Summary}
\label{sec:summary}
In this paper,
we have studied the properties of charmonium moving in a strongly coupled medium
from a holographic model. The holographic model in question reproduces the vacuum
properties of charmonium which makes it a suitable candidate for the current study.
Results for the spectral functions for both the transverse and longitudinal channels are reported in Figs.~\ref{fig:rhoT} and~\ref{fig:rhosub}.
Agreements with  lattice results \cite{Ding:2012pt} are observed.
Considering the on-going efforts on
understanding momentum-dependent spectral function from lattice,
our results from QCD-like strongly interacting thermal systems are timely.

Most importantly, we obtained the momentum dependence of the dissociation temperature $T_D(k)$ shown in Fig.~\ref{fig:Tdq}. This is determined by studying the momentum dependence of the height of the J/$\psi$ spectral peak.
It was observed that this dependence is associated with the residue of the J/$\psi$ pole rather than the width.
We also find that the momentum dependence of the spectral width is qualitatively consistent with
similar calculations from a perturbative QCD perspective.
The dispersion relations for J/$\psi$
have also been calculated and reveal no subluminal limiting velocity and show no indication of a peak in the
photon production rate.
Based on the dissociation temperature, we found that there would be no sharp J/$\psi$ peak in spectral density with momentum greater than $3.8 \sim 4.2$ GeV for temperatures relevant to the experiment.
Though it is too early to
translate the knowledge about $T_D(k)$ into experimental observables,
in particular,  the nuclear modification factor,
$R_{AA}$, the presence of a limiting momentum suggests a saturation of J/$\psi$ suppression
above some momentum.
To understand which properties of the spectral peak are the best indicator of $R_{AA}$,
simulations using phenomenological models
with a realistic representation of the fireball evolution,
such as kinetic rate equations, transport and statistical models
(see, for example, Ref.~\cite{Zhao:2010nk} and reference therein)
are required.
Holographic results presented here, can, in principle,
provide useful information to those simulations. This work has described many key features of charmonium
phenomenology and has demonstrated the usefulness of holographic models in describing them.

\acknowledgments
We would like to thank the Institute for Nuclear Theory at
the University of Washington for hospitality during the INT
Summer School on Applications of String Theory, when part of this
work was initiated.
We thank Tom Faulkner, Krishina Rajagopal for discussions on heavy quarks from ``top-down" holographic models,
Heng-Tong Ding for email communications on the lattice data
and Misha Stephanov for commenting on the draft.
Y.Y. would also like to thank Wai-Yee Keung,
Ho-ung Yee and Todd Springer for fruitful discussions.
The work of P.H was supported in part by US-NSF under Grant No. PHY-0969394, the work of Y.Y. in part by DOE under Grant No.\ DE-FG0201ER41195 and UIC dean's scholar fellowship.

\bibliography{finiteqcharm}
\end{document}